\documentclass[twoside,12pt]{article}
\usepackage{indentfirst}
\usepackage{bm}
\usepackage{cite}
\usepackage{graphicx}
\usepackage{epsfig}
\usepackage{amsmath}
\usepackage{amsfonts}
\usepackage{amssymb}
\usepackage{amsthm}
\usepackage{latexsym}
\usepackage{amsmath}

\usepackage{booktabs}
\usepackage{caption}
\newtheorem{thm}{Theorem}

\newtheorem{corollary}{Corollary}
\usepackage{epsfig}
\usepackage{amsmath}
\usepackage{epstopdf}
\usepackage{pgf,fancyhdr}
\usepackage{float}
\usepackage{comment}
\begin{document}

\begin{center}
\LARGE\bf Quantum Coherence and Basis-Dependent Correlations in Quantum Systems
\end{center}

\begin{center}
\rm  Zi-Han Chen$^{1}$, \ Hui Zhao$^{*1}$, \ Shao-Ming Fei$^{2}$, \ Zhi-Xi Wang$^{2}$, \ Mei-Ming Zhang$^{3}$
\end{center}

\begin{center}
\begin{footnotesize} \sl
$^1$ School of Mathematics, Statistics and Mechanics, Beijing University of Technology, Beijing 100124, China

$^2$  School of Mathematical Sciences, Capital Normal University, Beijing 100048, China

$^3$ School of Mathematics and Physics, Jiangsu University of Technology, Changzhou 213001, China

\end{footnotesize}
\end{center}
\footnotetext{\\{*} Corresponding author, Hui Zhao, zhaohui@bjut.edu.cn}
\vspace*{2mm}

\begin{center}
\begin{minipage}{15.5cm}
\parindent 20pt\footnotesize

Quantum coherence and quantum correlations are two fundamental features of quantum systems and are closely related in various quantum information tasks. In general, due to the existence of correlations, the amount of coherence in a bipartite state should be no less than that in the reduced states. We focus on what types of states have an equal amount of coherence in
the global state and the local states. For such states, the basis-dependent correlations (the difference between the coherence of a bipartite state and its reduced state with respect to a set of standard orthogonal local bases) vanish. Based on the Tsallis relative entropy coherence measure, we prove that block-diagonal product states exhibit vanishing basis-dependent correlations. Furthermore, we establish an equivalence between the absence of basis-dependent correlations and the block-diagonal product structure in a special case. We then consider the coherence measure based on quantum Fisher information and show that block-diagonal product states again lead to vanishing basis-dependent correlations, while in the two-qubit case the vanishing condition constrains the state to a broader block-diagonal-like product structure. These results provide a structural characterization of bipartite states with vanishing basis-dependent correlations under different coherence measures.
\end{minipage}
\end{center}

\begin{center}
\begin{minipage}{15.5cm}
\begin{minipage}[t]{2.3cm}{\bf Keywords:}\end{minipage}
\begin{minipage}[t]{13.1cm}
Coherence measure, Basis-dependent correlations, Tsallis relative entropy, Quantum Fisher information
\end{minipage}\par\vglue8pt
\end{minipage}
\end{center}

\section{Introduction}
Correlations and coherence originate from the superposition principle \cite{ref1}, leading to interference phenomena and non-classical behaviors \cite{ref2,ref3}. Correlations characterize statistics beyond the classical limit in composite quantum systems \cite{ref4}, including entanglement \cite{ref5} as well as more general non-classical forms such as discord \cite{ref6}, Bell nonlocality~\cite{ref7} and steering \cite{ref8}. Quantum coherence specifically quantifies the quantumness of a system by measuring how well superposition is preserved over time and space, serving as a foundational resource for quantum information tasks such as computing and sensing \cite{ref9}.

Generally, correlations do not depend on the choice of local bases, while coherence is basis dependent. To quantify quantum coherence, Baumgratz et al. \cite{ref10} proposed a rigorous framework for coherence measures such as the $l_1$-norm coherence and the relative entropy coherence. Yu and Song showed that the concurrence of bipartite pure states just captures the maximal difference between local coherence and the average coherence of one subsystem induced by local operations on the other subsystem with the assistance of classical communication \cite{ref11}. Chitambar and Hsieh unified the resource theories of entanglement and coherence by studying their behaviors in the operational setting of local incoherent operations and classical communication \cite{ref12}. Ref.~\cite{ref13}$-$Ref.~\cite{ref15} primarily investigated the relationship between quantum coherence and entanglement, including the characterization of entanglement through coherence measures based on R\'enyi relative entropies. Yu et al. divided the coherence of a quantum state into several classes and found a correspondence between geometric quantum discord and particular classes of quantum coherence \cite{ref16}. Ma et al. proved that the creation of quantum discord by multipartite incoherent operations is bounded by the amount of quantum coherence consumed in its subsystems during the process \cite{ref17}. Further, Ref.~\cite{ref18}$-$Ref.~\cite{ref20} investigated the relationship between partial coherence and quantum correlations.

A common strategy to eliminate the basis dependence of coherence is to take the extremal values \cite{ref6}. Yadin et al. defined basis-dependent correlations based on the differences of coherence between the bipartite state and its reduced states and presented the structure of the bipartite state when the basis-dependent correlation vanishes \cite{ref21}. Further, Li et al. provided a more concrete and intuitive representation of this structure, termed block-diagonal product states. Using the coherence measure based on skew information and the $l_1$-norm coherence, they derived the relationship between block-diagonal product states and states with vanishing correlations \cite{ref22}.

In this paper, we mainly focus on the coherence measures based on Tsallis relative entropy and on quantum Fisher information (QFI), and study the structure of bipartite states with vanishing basis-dependent correlations. The paper is organized as follows. In Section 2, we introduce relevant knowledge of the coherence measure based on the Tsallis relative entropy and the definition of block-diagonal product states, and discuss the relationship between block-diagonal product states and vanishing basis-dependent correlations. In Section 3, we generalize the definition of block-diagonal product states to block-diagonal-like product states, and discuss the relationship between block-diagonal-like product states, block-diagonal product states and vanishing basis-dependent correlations based on the QFI coherence measure. Conclusions are given in Section 4.

\section[Vanishing basis-dependent correlations using coherence based on Tsallis relative entropy]{Vanishing basis-dependent correlations\\
using coherence based on Tsallis relative entropy}
We first consider the relationship between the coherence measure and block-diagonal product states. Let $ \mathcal{H} $ be  a $ d $-dimensional Hilbert space with an orthonormal basis $ \{ |i \rangle \}_{i=1}^d $ and  $ \mathcal{S}(\mathcal{H}) $ the set of all quantum states on $ \mathcal{H}$. A state $ \rho \in \mathcal{S}(\mathcal{H}) $ is said to be incoherent
with respect to $ \{ |i \rangle \} $ if $\rho = \sum_{i=1}^d \lambda_i \, |i \rangle \langle i |$,
where $ \lambda_i \geq 0 $ and $ \sum_{i=1}^d \lambda_i = 1 $.
Denote by $\mathcal{I}(\mathcal{H})$ the set of all incoherent states on $\mathcal{H}$.

The orthonormal basis $ \{ |i \rangle \} $ defines the rank-1 projective measurement
$\Pi = \left\{ \, |i \rangle \langle i | \, \right\}_{i=1}^d$.
The coherence with respect to $ \{ |i \rangle \} $ is often identified with the coherence
induced by the von Neumann measurement $ \Pi $.
A L\"{u}ders measurement is specified by
$P = \{ P_k \}_{k=1}^m$, where $\{P_k \}$ is a set of orthogonal projectors on $ \mathcal{H} $ with $\operatorname{rank}\left( P_k \right) \geq 1$, $ P_k P_{k'} = 0 $ for all $ k \neq k' $, and $ \sum_{k=1}^m P_k = I $ with $ I $ being the identity operator on $ \mathcal{H} $.
Thus, $ \mathcal{H} $ can be decomposed as
$
\mathcal{H} = \bigoplus_{k=1}^m \mathcal{H}_k,
$
where $ \bigoplus $ denotes the direct sum of subspaces. We say that a state $ \rho $ is block incoherent if $ \rho = \sum_k P_k \rho P_k$.

Consider bipartite states $\rho^{AB}\in\mathcal{S}(\mathcal{H}^A\otimes\mathcal{H}^B)$, with $\dim\mathcal{H}^A = d_1$ and $\dim\mathcal{H}^B = d_2$, respectively.  The von Neumann measurement $\Pi^A = \{\Pi_i^A\}_{i=1}^{d_1}$ on $\mathcal{H}^A$, with $\Pi_i^A=|i\rangle\langle i|$, naturally induces a L\"{u}ders measurement $\Pi^A \otimes I^B = \{\Pi_i^A\otimes I^B\}_{i=1}^{d_1}$ on $\mathcal{H}^A \otimes \mathcal{H}^B$, where $ I^B $ denotes the identity operator on $ \mathcal{H}^B $.

In general, due to the  correlations in $\rho^{AB}$, the coherence $C\bigl(\rho^{AB},\, \Pi^A \otimes I^B\bigr)$ of $\rho^{AB}$ with respect to the measurement $\Pi^A \otimes I^B$ is no less than the coherence $C\bigl(\rho^A,\, \Pi^A\bigr)$ of its reduced state $\rho^A$ with respect to $\Pi^A$ for any coherence measure $C$, i.e.,
$
C\bigl(\rho^{AB},\, \Pi^A \otimes I^B\bigr) \;\geq\; C\bigl(\rho^A,\, \Pi^A\bigr).
$
The basis-dependent correlation of $\rho^{AB}$ with respect to $\Pi^A$ is defined by
\begin{equation}
D_C\bigl(\rho^{AB},\, \Pi^A\bigr) \;=\; C\bigl(\rho^{AB},\, \Pi^A \otimes I^B\bigr) \;-\; C\bigl(\rho^A,\, \Pi^A\bigr).
\label{eq1}
\end{equation}
Thus, $D_C\bigl(\rho^{AB},\, \Pi^A\bigr)$ quantifies the excess global coherence, with respect to the chosen local reference basis, beyond the coherence already contained in subsystem $A$.
As a simple case, if $\rho^{AB}$ is incoherent with respect to $\Pi^A \otimes I^B$, then $C\bigl(\rho^{AB},\, \Pi^A \otimes I^B\bigr) = C\bigl(\rho^A,\, \Pi^A\bigr)$, and hence $D_C\bigl(\rho^{AB},\, \Pi^A\bigr) = 0$.
Taking the relative entropy of coherence as the measure of coherence, Yadin et al. proved that $D_C\bigl(\rho^{AB},\, \Pi^A\bigr) = 0$ implies that
\begin{equation}
\rho^{AB} \;=\; \sum_k p_k \, \rho^A_k \otimes \rho^B_k,\label{eq2}
\end{equation}
where $\{\rho^A_k\}$ are states on $\mathcal{H}^A$ that are perfectly distinguishable by the measurement $\Pi^A$, and $\{p_k\}$ forms a valid probability distribution over $k$~\cite{ref21}.
Li et al. provided a more concrete and intuitive representation of  $\rho^{AB}$ in Eq.~\eqref{eq2}, termed a block-diagonal product state with respect to the measurement $\Pi^A$,
\begin{equation}
\rho^{AB} \;=\; \bigoplus_{k=1}^n p_k \, \rho_k^A \otimes \rho_k^B,
\label{eq3}
\end{equation}
where $\{P_k\}_{k=1}^n$ is a coarse-grained measurement of $\Pi^A$ with $P_k=\sum_{i\in I_k}\Pi_i^A$ and $ \{I_k\}$ a family of index sets satisfying $I_k \cap I_{k'} = \emptyset$ for $k \neq k'$ and $  \bigcup_{k=1}^n I_k = \{1, 2, \dots, d_1\}$, $\rho_k^A \in \mathcal{S}(P_k \mathcal{H}^A)$  and $\rho_k^B \in \mathcal{S}(\mathcal{H}^B)$, and $\{p_k\}$ forms a valid probability distribution.
Taking the skew information of coherence ($l_1$-norm coherence) as the coherence measure, they show that \cite{ref22}
$ D_C(\rho^{AB}, \Pi^A) = 0 $ iff  $\rho^{AB}$ is a block-diagonal product state with respect to the measurement $\Pi^A$.

In this section, we consider the coherence based on Tsallis relative entropy defined by \cite{ref23}
\begin{equation}
C_{T,\alpha}(\rho, P) \;=\; \frac{1}{\alpha \;-\; 1} \left\{ \sum_k \left[ \mathop{\text{tr}}\!\left( \left( P_k \rho^\alpha P_k \right)^{\frac{1}{\alpha}} \right) \right]^\alpha \;-\; 1 \right\},
\label{eq4}
\end{equation}
where $P = \{P_k\}$ is a L\"{u}ders measurement and \(\alpha \in (0,1) \cup (1,2]\).

\begin{thm}  \label{thm1}
Let $\Pi^A = \left\{ |i\rangle\langle i| \right\}_{i=1}^{d_1}$ be a von Neumann measurement  on $\mathcal{H}^A$. If $\rho^{AB}\in\mathcal{S}(\mathcal{H}^A \otimes \mathcal{H}^B)$ is a block-diagonal product state with respect to $\Pi^A$, then the basis-dependent correlation vanishes, i.e., $D_{C_{T,\alpha}}(\rho^{AB}, \Pi^A)=C_{T,\alpha}(\rho^{AB}, \Pi^A \otimes I^B)- C_{T,\alpha}(\rho^A, \Pi^A)=0$.
\end{thm}

\begin{proof}
A block-diagonal product state $\rho^{AB}$ with respect to $\Pi^A$ can be written as
\begin{equation*}
\rho^{AB} \;=\; \bigoplus_{k=1}^n p_k \, \rho_k^A \otimes \rho_k^B.
\label{eq17}
\end{equation*}
Because the supports corresponding to different $k$ are mutually orthogonal, the power of $\rho^{AB}$ preserves the direct sum structure,
\begin{equation}
(\rho^{AB})^{\alpha} \;=\; \bigoplus_{k=1}^n p_k^{\alpha} \, (\rho_k^A)^{\alpha} \otimes (\rho_k^B)^{\alpha}.
\label{eqq}
\end{equation}
Substituting Eq.~\eqref{eqq} into Eq.~\eqref{eq4} gives
\begin{equation}
C_{T,\alpha}(\rho^{AB}, \Pi^A \otimes I^B) = \frac{1}{\alpha - 1} \left\{
\sum_{k} p_k \sum_{i} \mathrm{tr}\left[ \Pi_i^A \, (\rho_k^A)^\alpha \, \Pi_i^A \otimes (\rho_k^B)^\alpha \right]^{\frac{1}{\alpha}} - 1\right\}.
\end{equation}
Using $(X\otimes Y)^{1/\alpha} =X^{1/\alpha}\otimes Y^{1/\alpha}$ for positive $X$ and $Y$, together with the normalization condition $\mathrm{tr} \rho_k^B =1$, we have
\begin{equation}
C_{T,\alpha}(\rho^{AB}, \Pi^A \otimes I^B) = \frac{1}{\alpha - 1} \left\{\sum_{k} p_k \sum_{i} \mathrm{tr}\left[(\Pi_i^A(\rho_k^A)^{\alpha}\Pi_i^A)^{\frac{1}{\alpha}}\right] - 1 \right\}.
\end{equation}

On the other hand, tracing out subsystem $B$ gives
$
(\rho^A)^\alpha \;=\; \sum_{k} p_k^\alpha \, (\rho_k^A)^\alpha.
$
Applying it to Eq.~\eqref{eq4}, we have
\begin{equation}
C_{T,\alpha}(\rho^{A}, \Pi^A ) = \frac{1}{\alpha - 1} \left\{\sum_{k} p_k \sum_{i} \mathrm{tr}\left[(\Pi_i^A(\rho_k^A)^{\alpha}\Pi_i^A)^{\frac{1}{\alpha}}\right] - 1 \right\}.
\label{eq23}
\end{equation}
Thus, $C_{T,\alpha}(\rho^{AB}, \Pi^A \otimes I^B) = C_{T,\alpha}(\rho^A, \Pi^A).$
That is,
$$
D_{C_{T,\alpha}}(\rho^{AB}, \Pi^A) = C_{T,\alpha}(\rho^{AB}, \Pi^A \otimes I^B) - C_{T,\alpha}(\rho^A, \Pi^A) = 0.
$$
\end{proof}

We next focus on the case $\alpha = \frac{1}{2}$, for which the vanishing of the basis-dependent correlation admits an exact structural characterization. In this case, the block-diagonal product structure is not only sufficient but also necessary. For $\alpha = \frac{1}{2}$, the coherence based on Tsallis relative entropy takes the form
\begin{equation}
C_{T,\frac{1}{2}}(\rho, P) \;=\; 2\left\{ \sum_k \mathop{\text{tr}}\!\left[ \left( P_k \sqrt{\rho} P_k \right)^2 \right] \;-\; 1 \right\},
\label{eq24}
\end{equation}
where $P = \{P_k\}$ is a L\"{u}ders measurement.

For a von Neumann measurement $\Pi^A = \left\{ |i\rangle\langle i| \right\}_{i=1}^{d_1}$ on $\mathcal{H}^A$, suppose that $D_{C_{T,\frac{1}{2}}}(\rho^{AB}, \Pi^A) = 0$.
By the definition of the basis-dependent correlation, we have $$C_{T,\frac{1}{2}}(\rho^{AB}, \Pi^A \otimes I^B) = C_{T,\frac{1}{2}}(\rho^A, \Pi^A),$$ which implies
\begin{equation}
\sum_i \mathop{\text{tr}}\!\left[ \left( (\Pi_i^A \otimes I^B) \sqrt{\rho^{AB}} (\Pi_i^A \otimes I^B) \right)^2 \right] = \sum_i \mathop{\text{tr}}\!\left[ \left( \Pi_i^A \sqrt{\rho^A} \Pi_i^A \right)^2 \right].
\label{eq26}
\end{equation}
Using the operator decomposition used in Ref.~\cite{ref22}, we can expand $\sqrt{\rho^{AB}}$ as
\begin{equation}
\sqrt{\rho^{AB}} = \sum_{m=1}^{d_2^2} A_m \otimes Y_m,
\label{eq27}
\end{equation}
where $ A_m $ are observables on $\mathcal{H}^A$, and $\{Y_m\}_{m=1}^{d_2^2}$ form an orthonormal basis for the real Hilbert space of self-adjoint operators acting on the $ d_2 $-dimensional Hilbert space $\mathcal{H}^B$. Consequently, $\rho^A= \mathop{\text{tr}}\!_B \rho^{AB}= \sum_m A_m^2$.
Substituting the above expressions into the zero-correlation condition and using the orthonormality of $\left\{ Y_m\right\}$, for each $i$, we obtain
\begin{equation}
\sum_m \mathop{\text{tr}}\!\left( A_m \Pi_i^A A_m \Pi_i^A \right) = \mathop{\text{tr}}\!\left( \sqrt{\sum_m A_m^2} \Pi_i^A  \sqrt{\sum_m A_m^2} \Pi_i^A \right).
\label{eq31}
\end{equation}
On the other hand, by Lieb's Theorem~\cite{ref24},
\begin{equation}
\mathop{\text{tr}}\!\left( \sqrt{\sum_m A_m^2} \Pi_i^A \sqrt{\sum_m A_m^2} \Pi_i^A \right) \ge \sum_m \mathop{\text{tr}}\!\left( |A_m| \Pi_i^A |A_m| \Pi_i^A \right),\; |A_m| = \sqrt{{A_m}^\dagger A_m}, \label{eq32}
\end{equation}
and using Jensen's inequality~\cite{ref25}, we get
\begin{equation}
\sum_m \mathop{\text{tr}}\!\left( |A_m| \Pi_i^A |A_m| \Pi_i^A \right) \ge \sum_m \mathop{\text{tr}}\!\left( A_m \Pi_i^A A_m \Pi_i^A \right). \label{eq33}
\end{equation}
Combining these inequalities with equality imposed by $D_{C_{T,\frac{1}{2}}}(\rho^{AB}, \Pi^A) = 0$, we find that both inequalities must be saturated. Hence,
\begin{equation}
\mathop{\text{tr}}\!\left( \sqrt{\sum_m A_m^2} \Pi_i^A \sqrt{\sum_m A_m^2} \Pi_i^A \right) = \sum_m \mathop{\text{tr}}\!\left( |A_m| \Pi_i^A |A_m| \Pi_i^A \right) = \sum_m \mathop{\text{tr}}\!\left( A_m \Pi_i^A A_m \Pi_i^A \right).  \label{eq34}
\end{equation}

Let $ \mu_m = \mathop{\text{tr}} A_m^2 $, $ \rho_m = A_m^2 / \mu_m $ and $ \rho^A = \sum_m \mu_m \rho_m $.
By the first equality in Eq.~\eqref{eq34}, $ A_m^2 $ can be expressed in the form of a direct sum \cite{ref22},
\begin{equation}
A_m^2 \;=\; \bigoplus_k \mu_{m} \, p_{mk}\,\sigma_k,  \label{eq35}
\end{equation}
where $ \{p_{mk}\} $ forms a probability distribution, and $ \sigma_k \in \mathcal{S}(P_k \mathcal{H}^A) $.
$\sigma_k$ is determined by the minimal direct sum decomposition of $\rho^A$ with respect to $\Pi^A$,
\begin{equation}
\rho^A \;=\; \bigoplus_k p_k \, \sigma_k.
\label{eq36}
\end{equation}

By the second equality in Eq.~\eqref{eq34}, from Proposition~2 in Ref.~\cite{ref22}, we have
 \begin{equation}
 \sqrt{\rho^{AB}} = \bigoplus_k \sqrt{\sigma_k} \otimes \left( \sum_m a_{mk} Y_m \right),
 \label{eq37}
 \end{equation}
 where $ a_{mk} = \sqrt{\mu_m p_{mk}} $.
 Then
\begin{equation}
\rho^{AB} = \bigoplus_k  \sigma_k \otimes \left( \sum_m a_{mk} Y_m \right)^2.
\label{eq38}
\end{equation}
Let $ p_k = \mathop{\text{tr}} \left( \sum_m a_{mk} Y_m \right)^2 $, $ \rho_k^B = \left( \sum_m a_{mk} Y_m \right)^2 / p_k $ and $ \rho_k^A = \sigma_k $. From Eq.~\eqref{eq38}, $ \rho^{AB} $ can be expressed as
\begin{equation}
\rho^{AB} \;=\; \bigoplus_k p_k \, \rho_k^A \otimes \rho_k^B,
\label{eq39}
\end{equation}
namely, $ \rho^{AB} $ is a block-diagonal product state with respect to the von Neumann measurement $ \Pi^A $, where
$\rho_k^A $ is supported on the block subspaces defined by $ \Pi^A $.
Therefore, we obtain the following theorem.

\begin{thm}
Let $ \Pi^A = \left\{|i\rangle\langle i|\right\}_{i=1}^{d_1}$ be a von Neumann measurement on $\mathcal{H}^A$. For $\alpha =\frac{1}{2} $,
$$
D_{C_{T,\frac{1}{2}}}(\rho^{AB}, \Pi^A) = 0
$$
if and only if $\rho^{AB}$ is a block-diagonal product state with respect to $\Pi^A$.
\end{thm}

\section{Vanishing basis-dependent correlations using the coherence measure based on the QFI}

The quantum Fisher information \cite{ref26,ref27} establishes the fundamental bound on the precision of estimating an unknown parameter and plays a crucial role in quantum metrology. In this section, we investigate the relationship between basis-dependent correlation differences and the block-diagonal product states with respect to $\Pi^A$ under the coherence measure related to the QFI defined by
\begin{equation}
F\left(\rho, \Pi\right) = \sum_{i} F\left( \rho , |i\rangle\langle i| \right),\label{eq40}
\end{equation}
where $F \left( \rho , A \right) = \frac{1}{4} \mathrm{tr}\left( \rho L^2 \right),$
A is an observable, and \(L\) is the symmetric logarithmic derivative associated with the measurement defined by
$i \left[ \rho , A \right] = \frac{1}{2} \left( L \rho + \rho L \right).$
Let $\rho = \sum_{l} \lambda_l |\phi_l\rangle\langle \phi_l |$ be the spectral decomposition of $\rho$, with $\lambda_l $ being the eigenvalues and $|\phi_l\rangle $ being the corresponding normalized eigenvectors. Then the QFI can be expressed as
\begin{equation}
F\left( \rho , A\right) = \sum_{ll'} \frac{ 2 \left( \lambda_l -\lambda_{l'}\right)^2}{\lambda_l + \lambda_{l'}} |\langle \phi_l | A | \phi_{l'} \rangle|^2.
\label{eq41}
\end{equation}
For a pure state \(\rho = |\nu \rangle\langle \nu|\), the quantum coherence $C_F $ based on the QFI is defined as
$
C_F\left(\rho, \Pi\right) = \sum_{i}\frac{1}{4} F\left( |\nu\rangle\langle\nu|, |i\rangle\langle i| \right).
$
It has been verified that $C_F$ is a coherence measure satisfying faithfulness, monotonicity, strong monotonicity and convexity~\cite{ref28}.
However, it has been shown that $C_F$ is not a coherence measure for mixed states, as it does not decrease under incoherent operations~\cite{ref29}.
Ref.~\cite{ref30} resolved this issue by introducing auxiliary incoherent operations and extended it to arbitrary states through a convex-roof construction.
Ref.~\cite{ref31} proposed two natural coherence measures based on the QFI without involving the specific quantum incoherent operations.
One is the direct generalization of the coherence measure via QFI for mixed states,
\begin{equation}
C_F^D\left( \rho,\Pi\right) = \sum_i \frac{1}{4} F\left( \rho, |i\rangle\langle i| \right).
\label{eq42}
\end{equation}
The convex-roof extension to a mixed state $\rho = \sum_l p_l |\varphi_l\rangle\langle \varphi_l|$ is defined as
\begin{equation}
C_F^C\left( \rho,\Pi\right) = \min_{\{p_l, |\varphi_l\rangle\}} \sum_l p_l C_F\left( |\varphi_l\rangle, \Pi\right).
\label{eq43}
\end{equation}
Both $ C_F^D\left(\rho,\Pi\right)$ and $C_F^C\left(\rho,\Pi\right)$ are shown to be valid coherence measures. Moreover,
for any state $\rho$ and von Neumann measurement $\Pi$, $C_F^D\left(\rho,\Pi\right) = C_F^C\left(\rho,\Pi\right)$.
Below we adopt $C_F^D$ as the coherence measure based on the QFI.

\begin{thm}
If $\rho^{AB}\in\mathcal{S}(\mathcal{H}^A \otimes \mathcal{H}^B)$ is a block-diagonal product state with respect to
a von Neumann measurement $\Pi^A = \left\{ |i\rangle\langle i| \right\}_{i = 1}^{d_1}$,
the basis-dependent correlation vanishes, i.e.,
$$
D_{C_{F}^D}\left(\rho^{AB}, \Pi^A\right) = C_{F}^D\left(\rho^{AB}, \Pi^A \otimes I^B\right) - C_{F}^D\left(\rho^{A}, \Pi^A\right) = 0.
$$
\end{thm}

 \begin{proof}
Since $\rho^{AB}$ is a block-diagonal product state with respect to $\Pi^A$, it can be expressed as
\begin{equation}
\rho^{AB} = \bigoplus_{k =1}^{n} p_k \, \rho_k^A \otimes \rho_k^B.
\label{eq44}
\end{equation}
Consider the spectral decompositions of $\rho_k^A$ and $\rho_k^B$,
\begin{equation}
\rho_k^A = \sum_l \lambda_{kl} |\phi_{kl}\rangle\langle\phi_{kl}|, \quad \rho_k^B = \sum_m \gamma_{km} |\psi_{km}\rangle\langle\psi_{km}|,
\label{eq45}
\end{equation}
where $\lambda _{kl}\ge 0$, $\gamma_{km}\ge 0$, $\sum_l\lambda _{kl}=1$, $\sum_m \gamma_{km}=1$, $|\phi_{kl}\rangle \in P_k\mathcal{H}^A$, $|\psi_{km}\rangle \in \mathcal{H}^B$.
Substituting Eq.~\eqref{eq45} into Eq.~\eqref{eq44}, we have
\begin{equation}
\rho^{AB} = \sum_{klm} p_k \lambda_{kl} \gamma_{km} |\phi_{kl}\rangle\langle\phi_{kl}| \otimes |\psi_{km}\rangle\langle \psi_{km}|.
\label{eq46}
\end{equation}
Set $q_t = p_k\lambda_{kl}\gamma_{km}$, where $t=(k,l,m)$, and $|\varphi_t\rangle = |\phi_{kl}\rangle \otimes |\psi_{km}\rangle$. Eq.~\eqref{eq46} shows that the eigenvalues of $\rho^{AB}$ are $q_t$, and its normalized eigenstates are $|\varphi_t\rangle$. Thus, $\rho^{AB} = \sum_t q_t |\varphi_t\rangle\langle\varphi_t|$.
From Eq.~\eqref{eq41}, we have
\begin{equation}
F\left( \rho^{AB}, \Pi^A \otimes I^B \right) = \sum_{tt'} \frac{2\left(q_t - q_{t'}\right)^2}{q_t + q_{t'}}|\langle \varphi_t| \Pi^A \otimes I^B|\varphi_{t'}\rangle|^2,
\label{eq48}
\end{equation}
\begin{equation}
\langle \varphi_t|\left(|i\rangle\langle i| \otimes I^B\right)|\varphi_{t'}\rangle =\langle \phi_{kl}| i\rangle\langle i|\phi_{k'l'}\rangle\delta_{kk'}\delta_{mm'} = \langle \phi_{kl}|i\rangle\langle i |\phi_{kl'}\rangle,
\label{eq49}
\end{equation}
\begin{equation}
q_t + q_{t'} = p_k \lambda_{kl} \gamma_{km} + p_{k'} \lambda_{k'l'} \gamma_{k'm'} = p_k \gamma_{km} \left( \lambda_{kl} + \lambda_{kl'}\right),
\label{eq50}
\end{equation}
\begin{equation}
q_t - q_{t'} = p_k \lambda_{kl} \gamma_{km} - p_{k'} \lambda_{k'l'} \gamma_{k'm'} = p_k \gamma_{km} \left( \lambda_{kl} - \lambda_{kl'}\right).
\label{eq51}
\end{equation}
Then, by incorporating Eqs.~\eqref{eq49}--\eqref{eq51} into Eq.~\eqref{eq48}, we have
\begin{equation}
C_F^D\left( \rho^{AB}, |i\rangle\langle i| \otimes I^B \right) = \sum_{km} \sum_{ll'} \frac{p_k^2\gamma_{km}^2\left(\lambda_{kl} - \lambda_{kl'}\right)^2}{2 p_k \gamma_{km} \left( \lambda_{kl} + \lambda_{kl'}\right)} |\langle \phi_{kl}| i\rangle |^2 |\langle i |\phi_{kl'}\rangle|^2.
\label{eq52}
\end{equation}
According to the definition of block-diagonal product states with respect to $\Pi^A$, for a fixed $i$, there must exist a $k$ such that $i \in I_k$, and hence $|\phi_{kl}\rangle = \sum_{j \in I_k} c_j |j\rangle$.
If $i' \notin I_k$, then $|i'\rangle \notin P_k\mathcal{H}^A$ and $|i'\rangle$ is orthogonal to all $|j\rangle$ ($j \in I_k$). In this case, $\langle i'|\phi_{kl}\rangle = \langle i' |\sum_{j \in I_k} c_j |j\rangle = \sum_{j \in I_k} c_j \langle i'|j\rangle  = 0$.
Therefore, $C_F^D\left(\rho^{AB}, \Pi^A \otimes I^B \right) \neq 0$ if and only if $i \in I_k$.
Equation~\eqref{eq52} is further simplified to
\begin{equation}
C_F^D\left( \rho^{AB}, |i\rangle\langle i| \otimes I^B \right) = p_k C_F^D\left(\rho_k^A, |i\rangle\langle i| \right).
\label{eq53}
\end{equation}
Therefore,
\begin{equation}
C_F^D\left(\rho^{AB}, \Pi^A \otimes I^B\right) = \sum_k \sum_{i\in I_k} p_k C_F^D\left(\rho_k^A, |i\rangle\langle i|\right).
\label{eq54}
\end{equation}

Next, we calculate $C_F^D\left(\rho^A,\Pi^A\right)$ in a similar way.
The reduced state $\rho^A$ has the spectral decomposition $\rho^A = \sum_{kl} p_k \lambda_{kl} |\phi_{kl}\rangle\langle\phi_{kl}|$. We have
\begin{equation}
C_F^D\left(\rho^A, |i\rangle\langle i|\right) = \sum_{kl} \sum_{k'l'} \frac{\left(p_k \lambda_{kl} - p_{k'}\lambda_{k'l'}\right)^2}{2\left(p_k \lambda_{kl} + p_{k'}\lambda_{k'l'}\right)} |\langle \phi_{kl}|i\rangle|^2|\langle\phi_{k'l'}|i\rangle|^2.
\label{eq55}
\end{equation}
Similarly, there exists a $k$ such that $i \in I_k$, and thus $ |\langle\phi_{kl}|i\rangle|$ is non-zero if and only if $k = k'$.
Hence,
\begin{equation}
C_F^D\left(\rho^A, |i\rangle\langle i|\right)
= p_k C_F^D\left(\rho_k^A, |i\rangle\langle i|\right).
\label{eq56}
\end{equation}
Then,
\begin{equation}
C_F^D\left(\rho^A, \Pi^A\right) = \sum_i C_F^D\left(\rho^A, |i\rangle\langle i|\right) = \sum_k \sum_{i\in I_k} p_k C_F^D\left(\rho_k^A, |i\rangle\langle i|\right).
\label{eq57}
\end{equation}
By comparing Eq.~\eqref{eq54} with Eq.~\eqref{eq57}, we finally obtain
$
C_F^D\left(\rho^{AB}, \Pi^A \otimes I^B\right) = C_F^D\left(\rho^A, \Pi^A\right).
$

\end{proof}

Generally, it is challenging to calculate the symmetric logarithmic derivative in the QFI-based coherence measure.
To analyze the state $\rho^{AB}$ more intuitively when the basis-dependent correlation vanishes, we consider the case of $d_1 = d_2 = 2$. In this case, $\Pi^A = \{|0\rangle\langle 0|, |1\rangle\langle  1|\}$.
According to the definition of the block-diagonal product state with respect to $\Pi^A$, $\rho^{AB} = \bigoplus_{k=1}^n p_k \rho_k^A \otimes \rho_k^B$, there are only two cases, either $n=1$ or $n=2$.
For $n=1$, $\rho^{AB} = \rho^A \otimes \rho^B$ is a product state.
For $n =2$, $\rho^{AB} = p_1 \rho_1^A \otimes \rho_1^B + p_2 \rho_2^A \otimes \rho_2^B = p_1 |0\rangle\langle 0| \otimes \rho_1^B + p_2 |1\rangle\langle 1|\otimes \rho_2^B$.
Thus, $\rho^{AB}$ is a block-diagonal matrix on $\mathcal{H}^A$ and has the following form
$$\begin{pmatrix}
\alpha_1 & \alpha_2 & 0 & 0 \\
\alpha_3 & \alpha_4 & 0 & 0 \\
0 & 0 & \beta_1 & \beta_2 \\
0 & 0 & \beta_3 & \beta_4
\end{pmatrix}.
$$
We call $\rho^{AB}$ a block-diagonal-like product state if it is of the following form
$$\rho^{AB} = \begin{pmatrix}
\alpha_1 & \alpha_2 & s_1^\dagger  & 0 \\
\alpha_3 & \alpha_4 & 0 & s_2^\dagger \\
s_1 & 0 & \beta_1 & \beta_2 \\
0 & s_2 & \beta_3 & \beta_4
\end{pmatrix}.
$$

\begin{thm}
With respect to a von Neumann measurement $\Pi^A = \{ |i\rangle\langle i|\}_{i=0}^{1}$ on $\mathcal{H}^A$, if a bipartite state $\rho^{AB}\in\mathcal{S}(\mathcal{H}^A \otimes \mathcal{H}^B)$ satisfies $D_{C_F^D}\left(\rho^{AB}, \Pi^A\right) = 0$, then $\rho^{AB}$ is a block-diagonal-like product state.
\end{thm}

\begin{proof}
Consider a general two-qubit state in the Bloch representation,
\begin{equation}
\rho^{AB} = \frac{1}{4}\left( I_2 \otimes I_2 + \vec{n} \cdot \vec{\sigma}^A + \vec{m} \cdot \vec{\sigma}^B + \sum_{i,j=1}^3 b_{ij}\sigma_i \otimes \sigma_j \right),
\end{equation}
where $\sigma_i^A = \sigma_i \otimes I_2$, $ \sigma_i^B = I_2 \otimes \sigma_i$, $\sigma_i$, $i=1,2,3$, are the standard Pauli matrices, $\vec{n} =(n_1,n_2,n_3)$, $\vec{m} =(m_1,m_2,m_3)$, and $n_i,m_i,b_{ij}$ are real numbers.

For the reduced state $\rho^A$, the definition of QFI gives
\begin{equation}
F\left(\rho^A,|x_{\pm}\rangle\langle x_{\pm}|\right) = \operatorname{tr} \rho^A L_{x_{\pm}}^2,
\label{eq58}
\end{equation}
where $L_{x_{\pm}}$ is the symmetric logarithmic derivative satisfying $i \left[ |x_{\pm}\rangle\langle x_{\pm}|, \rho^A\right] = -\frac{1}{2}\{L_{x_{\pm}}, \rho^A\}$.
Take $\{\frac{I_2}{\sqrt{2}}, \frac{\sigma_x}{\sqrt{2}}, \frac{\sigma_y}{\sqrt{2}},\frac{\sigma_z}{\sqrt{2}}\}$ as an orthonormal basis of $\mathcal{H}^A$. Since $L_{x_{\pm}}$ is Hermitian,  $|x_{\pm}\rangle\langle x_{\pm}|$
can be linearly expressed in terms of this basis. We have
$\left\{L_{x_+}, \rho^A\right\} = n_2 \sigma_z - n_3 \sigma_y$ and
$\left\{L_{x_-}, \rho^A\right\} = n_3 \sigma_y - n_2 \sigma_z$.
Assume $L_{x_{\pm}} = c_{1,\pm} \sigma_y + c_{2,\pm} \sigma_z$.
Then $L_{x_+} = - n_3 \sigma_y + n_2 \sigma_z$ and $L_{x_-} = n_3 \sigma_y - n_2 \sigma_z$.
Hence, the coherence measure $C_F^D$ with respect to $\sigma_x$ is given by
\begin{equation}
C_F^D \left(\rho^A, \sigma_x\right) =  \sum_{\pm} \frac{1}{4} \operatorname{tr}\rho^A L_{x_{\pm}}^2 = \frac{1}{2} \left( n_2^2 + n_3^2 \right).
\label{eq64}
\end{equation}
Similarly, we have
\begin{equation}
C_F^D \left(\rho^A, \sigma_y\right) = \frac{1}{2} \left( n_1^2 + n_3^2 \right),
\label{eq65}
\end{equation}
\begin{equation}
C_F^D \left(\rho^A, \sigma_z\right) = \frac{1}{2} \left( n_1^2 + n_2^2 \right).
\label{eq66}
\end{equation}
Let us first calculate $F\left(\rho^{AB},\sigma_z \otimes I_2\right)$.
The eigenstates of $\sigma_z \otimes I_2$ are $\{|z_i \rangle\langle z_i|\}_{i=1}^4$.
According to the definition of QFI,
\begin{equation}
F\left( \rho^{AB}, |z_i\rangle\langle z_i| \right) = \operatorname{tr}\left(\rho^{AB}L_{z_i}^2\right),
\label{eq68}
\end{equation}
where $L_{z_i}$ is the symmetric logarithmic derivative satisfying
\begin{equation}
i\left[ \rho^{AB},  |z_i\rangle\langle z_i| \right] = \frac{1}{2} \left\{ L_{z_i}, \rho^{AB} \right\}.
\label{eq69}
\end{equation}
Selecting $\{\frac{\sigma_x \otimes I_2}{2}, \frac{I_2 \otimes \sigma_i}{2}, \frac{\sigma_i \otimes \sigma_j}{2}\}\left( i,j = x,y,z\right)$ as an orthonormal basis on $\mathcal{H}^A \otimes \mathcal{H}^B$, we can express
$L_{z_i}$ as
\begin{align}
| z_1\rangle\langle z_1| &= \frac{1}{4} I_2 \otimes I_2 + \frac{1}{4} I_2 \otimes \sigma_z + \frac{1}{4} \sigma_z \otimes I_2 + \frac{1}{4} \sigma_z \otimes \sigma_z,\\
| z_2\rangle\langle z_2| &= \frac{1}{4} I_2 \otimes I_2 - \frac{1}{4} I_2 \otimes \sigma_z + \frac{1}{4} \sigma_z \otimes I_2 - \frac{1}{4} \sigma_z \otimes \sigma_z,\\
| z_3\rangle\langle z_3| &= \frac{1}{4} I_2 \otimes I_2 + \frac{1}{4} I_2 \otimes \sigma_z - \frac{1}{4} \sigma_z \otimes I_2 - \frac{1}{4} \sigma_z \otimes \sigma_z,\\
| z_4\rangle\langle z_4| &= \frac{1}{4} I_2 \otimes I_2 - \frac{1}{4} I_2 \otimes \sigma_z - \frac{1}{4} \sigma_z \otimes I_2 + \frac{1}{4} \sigma_z \otimes \sigma_z.
\end{align}
Then,
\begin{equation}
C_F^D \left( \rho^{AB}, \sigma_z \otimes I_2\right) = \frac{1}{2} \left( n_1^2 + n_2^2 + m_1^2 + m_2^2 + b_{11}^2 +b_{12}^2 +b_{13}^2 + b_{21}^2 + b_{22}^2 + b_{23}^2 + b_{31}^2 + b_{32}^2 \right).
\label{eq83}
\end{equation}
According to the definition of $C_F^D$, we obtain
\begin{equation}
\begin{split}
C_F^D \left( \rho^{AB},  \Pi^A \otimes I^B \right) &= C_F^D \left( \rho^{AB}, |0\rangle\langle 0| \otimes I_2 \right) + C_F^D \left( \rho^{AB}, |1\rangle\langle 1| \otimes I_2 \right) \\
&= C_F^D \left( \rho^{AB}, \sigma_z \otimes I_2\right),
\end{split}
\label{eq86}
\end{equation}
\begin{equation}
\begin{split}
C_F^D \left( \rho^{A},  \Pi^A  \right) &= C_F^D \left( \rho^{A}, |0\rangle\langle 0| \right) + C_F^D \left( \rho^{A}, |1\rangle\langle 1| \right) \\
&= C_F^D \left( \rho^{A}, \sigma_z \right),
\end{split}
\label{eq87}
\end{equation}
where $\{|z_i \rangle\langle z_i|\}_{i=1}^4$ are the eigenstates corresponding to $\sigma_z \otimes I_2$, and $\{| z_\pm\rangle\langle z_\pm |\}$ are the eigenstates corresponding to $\sigma_z $.

Taking into account Eq.~\eqref{eq66}, Eq.~\eqref{eq83}, Eq.~\eqref{eq86} and Eq.~\eqref{eq87}, we obtain
\begin{equation}
\frac{1}{2} \left( n_1^2 + n_2^2 + m_1^2 + m_2^2 + b_{11}^2 +b_{12}^2 +b_{13}^2 + b_{21}^2 + b_{22}^2 + b_{23}^2 + b_{31}^2 + b_{32}^2 \right) = \frac{1}{2} \left( n_1^2 + n_2^2 \right).
\label{eq88}
\end{equation}
Since each term in Eq.~\eqref{eq88} is a real number, we have
\begin{equation}
m_1^2 = m_2^2 = b_{11}^2 = b_{12}^2 = b_{13}^2  = b_{21}^2 = b_{22}^2 = b_{23}^2 = b_{31}^2 = b_{32}^2 = 0.
\label{eq89}
\end{equation}
Then $\rho^{AB}$ can be written as
$$
\rho^{AB} = \begin{pmatrix}
\frac{1 + n_3 + m_3 + b_{33}}{4} & 0 & \frac{n_1 - in_2}{4} & 0 \\
0 & \frac{1+n_3-m_3-b_{33}}{4} & 0 & \frac{n_1 - in_2}{4} \\
\frac{n_1 + in_2}{4} & 0 & \frac{1 - n_3 + m_3 - b_{33}}{4} & 0 \\
0 & \frac{n_1 + in_2}{4} & 0 & \frac{1 - n_3 - m_3 + b_{33}}{4}
\end{pmatrix},
$$
namely, $\rho^{AB}$ is a block-diagonal-like product state.
\end{proof}

Theorem 4 shows that vanishing basis-dependent correlation restricts a two-qubit state to the block-diagonal-like class. A block-diagonal product state is recovered when the remaining off-diagonal terms connecting the two reference subspaces of subsystem $A$ also vanish. This immediately gives the following corollary.

\begin{corollary}
If $D_{C_F^D}\left(\rho^{AB}, \Pi^A\right) = 0$ and $n_1 = n_2=0$, then $\rho^{AB}$ is a block-diagonal product state with respect to the von Neumann measurement $\Pi^A = \{ |i\rangle\langle i|\}_{i=0}^{1}$ on $\mathcal{H}^A$.
\end{corollary}

\section{Conclusion}
We have investigated the relationship between quantum coherence and basis-dependent correlations in bipartite quantum systems from the perspective of different coherence measures. For the coherence measure based on the Tsallis relative entropy, we have shown that block-diagonal product states exhibit vanishing basis-dependent correlations. In particular, for $\alpha=1/2$, the vanishing condition provides a complete characterization of the block-diagonal product structure. We have further considered coherence based on QFI and found that block-diagonal product states also lead to vanishing basis-dependent correlations. For two-qubit systems, the absence of such correlations imposes a broader block-diagonal-like structure on the state. These results reveal that vanishing basis-dependent correlations are closely related to specific structural properties of bipartite states, while the resulting constraints depend on the coherence measure employed. Our results provide a further perspective on the interplay between quantum coherence, correlations, and the structure of composite quantum states.

\noindent\textbf{CRediT authorship contribution statement}

{\bf Zi-Han Chen, Hui Zhao:} Formal analysis, Writing, Calculation and figure.

{\bf Shao-Ming Fei, Zhi-Xi Wang, Mei-Ming Zhang:} Writing, review and editing.\\

\noindent\textbf{Declaration of competing interest}

The authors declare no conflicts of interest.\\

\noindent\textbf{Data availability}

No data were used for the research described in the article.\\

\noindent\textbf {Acknowledgements}
This work was supported by Beijing Key Laboratory of AI Technology for Financial Payment Supervision under Grant No.~BZ-2026-044, the National Key R\&D Program of China under Grant No.~2022YFB38\allowbreak06000, the National Natural Science Foundation of China under Grants (12272011, 12075159, 12126351 and 12171044), and the Academician Innovation Platform of Hainan Province.

\renewcommand{\theequation}{A\arabic{equation}}
\setcounter{equation}{0}

\end{document}